\documentclass{article}
\usepackage[T1]{fontenc}
\usepackage{wrapfig}
\usepackage[portuges,english]{babel}
\usepackage{amssymb,latexsym,amsmath,color,mathrsfs,ifsym,graphics,stmaryrd} 
\usepackage[colorlinks,linkcolor=blue,urlcolor=blue,citecolor=black,
plainpages=false,pdfpagelabels,breaklinks]{hyperref}
\usepackage{dialogue}
\usepackage{graphicx}
\title{{\sc A Realist Approach to Quantum Individuality:\\Against the ``Received'' and ``Alternative'' Views}}

\author{{\sc Christian de Ronde}\thanks{Fellow Researcher of the Consejo Nacional de Investigaciones Cient\'{\i}ficas y T\'ecnicas. E-mail: cderonde@gmail.com}\ $^{1}$$^{2}$$^{3}$ \ and {\sc Ivan Klarreich$^{4}$}}
\date{}

\usepackage[margin=2.15cm]{geometry}

\begin{document}
\maketitle
\begin{center}
\begin{small}
1. Philosophy Institute Dr. A. Korn, Buenos Aires University - CONICET\\ 
2. Engineering Institute - National University Arturo Jauretche, Argentina.\\
3. Center Leo Apostel for Interdisciplinary Studies, Brussels Free University, Belgium.\\
4. Faculty of Philosophy and Letters, Buenos Aires University.
\end{small}
\end{center}

\bigskip 

\begin{abstract}
\noindent In this work we address the contemporary debate about quantum individuality expressed as an opposition between those who, like D\'ecio Krause, defend the reference of the theory to ``non-individual particles'', and those like Dennis Dieks, who propose instead to preserve the notion of ``classical particle'' as commonly applied by contemporary physicists. We will argue that these viewpoints rather than truly opposed share a common methodology which has helped to reinforce the doctrine of classical concepts that was imposed by Bohr and Dirac within the ``standard'' formulation  ---which remains the orthodox physical account of the theory of quanta. We will also discuss a recently proposed relativist yet objective relational account of quantum individuality \cite{deRondeFMMassri24} which, going back to Einstein's methodological approach, opens the doors to a completely new, truly realist understanding of quantum individuality.  
\\

\end{abstract}

\renewenvironment{enumerate}{\begin{list}{}{\rm \labelwidth 0mm
\leftmargin 0mm}} {\end{list}}

\newcommand{\ita}{\textit}
\newcommand{\mcal}{\mathcal}
\newcommand{\mfrak}{\mathfrak}
\newcommand{\mbb}{\mathbb}
\newcommand{\mrm}{\mathrm}
\newcommand{\msf}{\mathsf}
\newcommand{\mscr}{\mathscr}
\newcommand{\lra}{\leftrightarrow}
\renewenvironment{enumerate}{\begin{list}{}{\rm \labelwidth 0mm
\leftmargin 5mm}} {\end{list}}

\newtheorem{theo}{Theorem}[section]
\newtheorem{definition}[theo]{Definition}
\newtheorem{lem}[theo]{Lemma}
\newtheorem{met}[theo]{Method}
\newtheorem{prop}[theo]{Proposition}
\newtheorem{coro}[theo]{Corollary}
\newtheorem{exam}[theo]{Example}
\newtheorem{rema}[theo]{Remark}{\hspace*{4mm}}
\newtheorem{example}[theo]{Example}
\newcommand{\ninv}{\mathord{\sim}} 
\newtheorem{postulate}[theo]{Postulate}
\newcommand{\Proof}{\textit{Proof:} \,}
\newcommand{\cqd}{\hfill{\rule{.70ex}{2ex}} \medskip}

\bigskip

\bigskip

\begin{flushright}
{\small {\it Everything is made of atoms.}\\
\smallskip
Richard Feynman} 
\end{flushright}


\section*{Introduction}

Today, even though there is no agreement about the general understanding of the theory of quanta, there is ---anyhow--- an established consensus in both physics and philosophy of science regarding its {\it reference}. Indeed, according to physicists and philosophers alike, Quantum Mechanics (QM) describes a microscopic realm constituted by elementary quantum particles such as electrons, protons and quarks. As explained by the U.S. physicists Richard Feynman \cite[Chap. 37]{Feynman63}: ``Quantum mechanics is the description of the behavior of matter and light in all its details and, in particular, of the happenings on an atomic scale.'' Yet, strange as it might seem, this statement is in conflict with another famous statement made by Feynman \cite[p. 129]{Feynman67} himself,  according to which: ``nobody understands quantum mechanics.'' And indeed, as exposed in a recent article by Natalie Wolchover titled: {\it What Is a Particle?}, the contemporary physics community seems unable to clearly explain nor even agree on what are these unobservable entities. Instead, due to the multiplication of models, the notion of ``quantum particle'' has fragmented itself in completely different incompatible directions as, for example: the collapse of a quantum wave function, the excitation of a quantum field, the irreducible representation of a group, vibrating strings, the deformation of a qubit ocean, measurement outcomes, etc. (see for a detailed analysis \cite{Wolchover20}). Unfortunately, the situation is not any different in the philosophical arena, where the exponential multiplication of ``interpretations'' ---as philosophers call their narratives--- has added a new level of fragmentation through the introduction of potentialities, propensities, flashes, quantum fields, consciousness, multiverses, etc. And yet, regardless of the many difficulties, the atomist discourse has remained a dominant ``way of talking'' about quanta that underlies contemporary physical and philosophical research.


As it is well known, the reference to ``microscopic particles'' was famously established during the early 1930s by the ``Standard'' formulation of QM (SQM) as designed by one of the most influential figures of the 20th century, the Danish physicist Niels Bohr. Already in 1913 Bohr presented his famous quantum model of the atom as an essentially inconsistent set of rules effectively supplemented by a narrative about ``small planetary systems''. The lack of consistency between his model and classical physics would become bridged through the introduction of a {\it correspondence principle} that, according to Arnold Sommerfeld \cite{BokulichBokulich20}, would allow Bohr to magically connect the classical and the quantum realms: ``Bohr has discovered in his principle of correspondence a magic wand (which he himself calls a formal principle), which allows us immediately to make use of the results of the classical wave theory in the quantum theory.'' In turn, the Danish physicist \cite[p. 313]{Bohr48} would impose his famous {\it doctrine of classical concepts} according to which: ``However far quantum effects transcend the scope of classical physical analysis, the account of the experimental arrangement and the record of the observations must always be expressed in common language supplemented with the terminology of classical physics.'' In this way, the complementary reference to `waves' and `particles' would become a limit constraining physical representation itself: 
\begin{quotation}
\noindent {\small ``According to the view of the author, it would be a misconception to believe that the difficulties of the atomic theory may be evaded by eventually replacing the concepts of classical physics by new conceptual forms. [...] No more is it likely that the fundamental concepts of the classical theories will ever become superfluous for the description of physical experience. The recognition of the indivisibility of the quantum of action, and the determination of its magnitude, not only depend on an analysis of measurements based on classical concepts, but it continues to be the application of these concepts alone that makes it possible to relate the symbolism of the quantum theory to the data of experience.'' \cite[p. 16]{Bohr29}} 
\end{quotation}
Bohr's {\it complementarity principle}, understood as justifying the use of contradictory representations to account for quantum entities, would remain a methodological guide within both physics and philosophy. For example, in the third volume of the {\it Feynman's Lectures on Physics} \cite[Chap. 37]{Feynman63}, the U.S. physicist would famously argue that the double-slit experiment ``contains the only mystery of quantum mechanics''. As a consequence, the self-contradictory reference to `waves' and `corpuscles' has remained part of the mainstream contemporary discourse used sometimes to ``describe microscopic reality'' and sometimes ``just as a way of talking''. 

After the dispute that had taken place between Heisenberg's matrix formulation and Schr\"odinger's wave mechanics, Bohr's main ideas would become re-introduced within the theory through an axiomatization presented by a young English engineer and mathematician \cite{Dirac74}. Dirac's proposal, staying close to Bohrian and positivist premises, would become finally established during the early 1930s as the orthodox ``standard'' account of the theory of quanta. Following Bohr's reference to ``small particles'', Dirac \cite[p. 1]{Dirac74} would argue explicitly that: ``it has been found possible to set up a new scheme, called quantum mechanics, which is more suitable for the description of phenomena on the atomic scale.'' And ---also--- following Bohr's complementarity discourse: ``We have, on the one hand, the phenomena of interference and diffraction, which can be explained only on the basis of a wave theory; on the other, phenomena such as photo-electric emission and scattering by free electrons, which show that light is composed of small particles.'' According to the English engineer and mathematician: ``quantum mechanics is able to effect a reconciliation of the wave and corpuscular properties of light''  such that ``[a]ll kinds of particles are associated with waves in this way and conversely all wave motion is associated with particles. Thus all particles can be made to exhibit interference effects and all wave motion has its energy in the form of quanta.'' 

The solitary resistance exerted  during the first decades of the 20th century by Albert Einstein and Erwin Schr\"odinger regarding the paradoxical reference of SQM to physical reality would be finally crushed in 1935 when the famous EPR confrontation would be settled in favor of Bohr. As described by Jim Baggott \cite{Baggott21}: ``The popular reading of subsequent history suggests that Bohr emerged the victor in the debate, browbeating the presumed-senile Einstein into submission, and the Copenhagen interpretation became a dogmatic orthodoxy.'' After the IIWW the physics community would soon learn to accept not only an inconsistent ---partly linear and partly nonlinear--- axiomatic formalism but also an even more inconsistent narrative about corpuscles, waves, contextual states, measurements and collapses. Physicists would then claim that while SQM successfully explained the microscopic foundation of matter, no one really  understood what the theory was talking about. Within this instrumentalist turn, the foundational and philosophical debates about QM would become silenced and ---even--- persecuted (e.g., \cite{Clauser03}). As described by Olival Freire Jr.: 
\begin{quotation}
\noindent {\small ``In the US, which after the Second World War became the central stage of research in physics in the West, the discussions about the interpretation of quantum mechanics had never been very popular. A common academic policy was to gather theoreticians and experimentalists in order to favour experiments and concrete applications, rather than abstract speculations (Schweber 1986). This practical attitude was further increased by the impressive development of physics between the 1930s and the 1950s, driven on the one hand by the need to apply the new quantum theory to a wide range of atomic and subatomic phenomena, and on the other hand by the pursuit of military goals. As pointed out by Kaiser (2002, pp. 154-156), `the pedagogical requirements entailed by the sudden exponential growth in graduate student numbers during the cold war reinforced a particular instrumentalist approach to physics'.'' \cite[pp. 77-78]{Freire15}} 
\end{quotation} 

However, against all odds, a few decades later, during the mid-1960s, John Bell would return in secrecy to the forgotten works of Einstein. ``On leave of absence from SLAC and CERN''\footnote{A warning added to the manuscript that made clear the dangers of discussing the works of Einstein.}  Bell would write a paper titled, ``On the Einstein Podolsky Rosen Paradox'' that, once again, would light the spark of the realist debates that physicists had learn to forget \cite{Bell64}. Some years later, Bell's paper would find John Clauser, a young U.S. experimental physicist who would finally test Einstein's hypothesis regarding entanglement. One decade later, in 1981, Alain Aspect and his group would confirm Clauser's results giving momentum to a rising field of research called ``philosophy of quantum mechanics'' where, once again, the debate about the meaning and reference of the theory would become publicly addressed. It is in this context that not only physicists but also philosophers, mathematicians and logicians would engage in one of the most fascinating debates regarding the meaning of QM.\footnote{A good example are the famous meetings in Johensu, Finland, in the years 1985, 1987 and 1990, organized by the physicist Kalervo Laurikainen.} In particular, one of the problems that would become commonly discussed within this new field of philosophical expertise is the problem of representing the ``quantum particles'' physicists repeatedly talked about. 

Today, after some decades of philosophical research, there are two main lines of interpretation regarding the problem of quantum individuality. The first, known as ``the Received View of quantum non-individuals'', has been defended by Newton da Costa, D\'ecio Krause, Ot\'avio Bueno and Jonas Arenhart ---between many others. According to this interpretation quantum particles must be understood as non-individuals. This position has been disputed by an ``Alternative View'' defended by Dennis Dieks, Andrea Lubberdink and Tomasz Bigaj ---between others. Instead of talking about non-individuals this second approach defends the idea that the classical notion of individual can be still considered, at least in certain regimes and depending on the context, as applicable within the theory of quanta. In this work we begin by critically addressing the common atomist ground of both ``received'' and ``alternative'' views. We will then continue to address a new proposal which departs the orthodox interpretational methodology and presents quantum individuality as necessarily constrained by two general principles of physical representation, namely, operational invariance and conceptual objectivity. We end this work with an analysis of the methodological dispute between the Bohrian and Einstenian approaches to physical representation itself.

\section{Quantum Individuality: The ``Received'' and ``Alternative'' Views}

The “Received View” (RV) on quantum non-individuality is an approach to the nature of quantum entities attributed by members of the Brazilian School in the Philosophy of Physics to Erwin Schrödinger.  Despite his decisive role in the initial development of quantum mechanics, the Austrian physicist —as well as Einstein— would express numerous critical reflections about the theory. Some of these, many years later, would become the cornerstone of the non-individuality approach. In a passage repeatedly quoted by RV proponents, Schrödinger claims:
\begin{quotation}
\noindent {\small “[...] we have yet been compelled to dismiss the idea that such a particle [the quantum particle] is an individual entity which retains its ‘sameness’ forever. Quite the contrary, we are now obliged to assert that the ultimate constituents of matter have no ‘sameness’ at all [...] It is not a question of our being able to ascertain the identity in some instances and not being able to do so in others. It is beyond doubt that the question of ‘sameness’, of identity, really and truly has no meaning.” \cite[p. 121-122]{Schrodinger14}}   
\end{quotation}

According to quantum theory, physical systems obey Fermi-Dirac or Einstein-Bose statistics in which, unlike classical statistics, the exchange or {\it permutation} of one entity for another of the same type is not considered in the counting of states. This characteristic that quantum theory attributes to the entities to which it refers prompted the idea that ``quantum particles'' of the same type share all their physical properties, that is, they are {\it indistinguishable}\footnote{In what follows, we will use the notions of `indistinguishability' and `indiscernibility' as synonyms.} from each other. Thus, according to the RV's reading of Schrödinger's writings, identity “fails” in QM. Since the 1990s, Décio Krause and various collaborators such as Steven French and Jonas Arenhart have developed —extending Newton da Costa's \cite{daCosta94} proposals in the field of logic and the metaphysics of science— an approach according to which quantum entities can be understood as \textit{non-individuals} \cite{FrenchKrause2006, Krause92}. Let us unpack this. The notion of non-individual is defined as a negation of the classical notion of individual, typically based on essential properties, epistemic conditions of re-identification, and the logical definition of identity. In this sense, Krause's approach to non-individuals can be dissected on three levels (see \cite{KrauseArenhart18}). At the metaphysical level, individuality is understood as the possession of a special property such as an essence or {\it haecceity} that makes an entity unique. A non-individual, therefore, is an entity that lacks essential properties of this kind. At the epistemic level, an individual is defined by the satisfaction of two conditions: an individual (i) belongs to a class in the sense that it is one of a class and (ii) can be re-identified as such in different contexts. Consequently, a non-individual is an entity that does not satisfy conditions (i) or (ii). Finally, at the logical-mathematical level, while an individual is one that conforms to the logical law of identity: $\forall x (x=x)$; a non-individual is one that does not conform to it. The main area of development in RV is the logical-mathematical level, in which a formal apparatus known as Quasi-Set Theory (QST) has been proposed with the aim of accounting for quantum entities as non-individuals. As maintained by Krause, QST would allow us to model the non-individual nature of quantum entities at the formal level by breaking the classical relationship between identity and indistinguishability. According to Leibniz's Law, indistinguishable objects are identical and identical objects are indistinguishable. In QST, by contrast, indistinguishability does not necessarily imply identity, i.e, the Principle of Identity of Indiscernibles (PII) is not generally valid. QST is, in short, a generalization of Zermelo-Fraenkel Set Theory in which identity does not apply to all objects as a result of the restriction of the logical law of identity.\footnote{Krause calls logical-mathematical devices that, like QST, suspend the generalized validity of identity, “non-reflexive.” However, it is important to distinguish the non-reflexivity to which Krause refers from a strict logical version of the notion. Since the notion of logical consequence underlying QST is classical (from which it follows that every fragment of the language entails itself, i.e., ‘A’ entails ‘A’) no QST formalism is non-reflexive in the sense that it rejects any of the Tarskian properties that define the logical consequence relation for all sentences in the system —for example, reflexivity. In the context of the discussion of individuality, the non-reflexivity to which Krause refers consists in the fact that the classical laws of identity —particular statements formulated in the language of the theory— hold for some objects and do not hold for others in apparatus such as QST (See \cite[Chap. 8]{FrenchKrause2006}).} Krause's strategy is simple. QST is a set theory with {\it Ur}-elements (ZFU) of two types: 
\begin{quotation}
\noindent {\small ``In the first category we have the {\it M}-atoms, which can be thought of as the macroscopic objects of our environment. They will be treated as Urelemente of ZFU stricto sensu; hence, we will admit that classical logic is valid with respect to them in all its aspects. The atoms of the other kind ({\it m}-atoms) may be intuitively thought of as elementary particles of modern physics, and we will suppose, following Schrödinger's ideas, that identity is meaningless with respect to them.'' \cite[p. 402]{Krause92}}   
\end{quotation}

The rupture of the classical relationship between identity and indistinguishability, then, consists of defining  the relationship of {\it indistinguishability} ($\equiv$) —and not identity— as the relationship that can be applied to any type of object in QST. Indistinguishability is defined by attributing to it the behavior typically associated with identity: reflexivity, symmetry, transitivity, and, with restrictions regarding the type of object, substitution. After that, classical identity (=) is defined by restricting its validity to \textit{M}-objects, that is, to macroscopic objects: $x=y ={_d}{_f}   \neg m (x) \wedge \neg m (y) \wedge x \equiv y$.  Intuitively, two objects are identical if they are indistinguishable and neither of them is microscopic. As French and Krause argue:
\begin{quotation}
\noindent {\small ``The basic idea is that the {\it M}-atoms have the properties of standard {\it Ur}-elemente of ZFU, while the {\it m}-atoms may be thought of as representing the elementary basic entities of quantum physics. With regard to these, the concept of identity does not apply. In quasi-set theory, this is achieved by restricting the concept of formula: expressions like x = y are not well formed if x and y denote \textit{m}-atoms. The equality symbol is not a primitive logical symbol, but a concept of {\it extensional identity} (represented by =${_E}$) is introduced by definition so that it has all the properties of standard identity, similar to the corresponding concept in ZFU. Thus, the axiomatics allows us to differentiate between the concepts of (extensional) {\it identity} (being the very same object) and {\it indistinguishability} (agreement with respect to all the attributes).'' \cite[p. 276]{FrenchKrause2006}}     
\end{quotation}

Thus, QST deals with two types of formal objects. On the one hand, classic elements of set theory, which conform to standard logical laws. On the other hand, QST models \textit{m}-objects, which do not conform to the logical law of identity and, furthermore, can be quantified but not ordered or labeled, that is, they have a defined cardinality, but cannot be assigned an ordinal. Given that quantum entities, considered as particles, are indistinguishable from one another, Krause concludes that quantum (non-)individuals cannot be identified by a formal label that functions as a rigid designator. According to this, we can know how many particles of the same type there are in a quantum system, but we cannot distinguish them from one another. In this sense, classical set theory (ZFU) is a fragment of quasi-set theories such as S* or $\mathfrak{Q}$, insofar as these include the classical treatment of ZFU objects but extend it to a more general theory that deals with collections of indistinguishable but not identical objects.

During the first decade of the 21st century, in opposition to the RV, Simon Saunders, Fred Muller and Michael Seevinck proposed a different approach to quantum individuality based on {\it weak discernibility} \cite{MullerSaunders08, MullerSeevinck09, Saunders06}. The authors argued that quantum entities, both elementary particles and quantum fields, are ---in certain circumstances--- discernible and PII can be applied to them. According to their characterization of weak discernibility, two particles {\it 1} and {\it 2} are weakly o {\it relationally} discernible (PII-R) if they are not discernible by physical properties —that is, they are not {\it absolutely} discernible (PII-A)— but they are discernible by some admissible physical relation in each state of the composite system of which they are part. Thus, quantum particles of the same type can be (weakly) discernible when they are in an {\it irreflexive but symmetric} relationship. From a physical point of view, an irreflexive relationship is not a type of relationship that an entity can have with itself, from which it follows that such a relationship links two entities, and not one and the same. For example, if we consider the case of fermions, represented by the state $\frac{1}{\sqrt{2}}$ \{$|\uparrow\rangle_1 |\downarrow\rangle_2$ — $|\downarrow\rangle_1 |\uparrow\rangle_2$\} the irreflexive but symmetric relationship is to have opposite spin directions. Dennis Dieks, among others, has criticized this approach for assuming that the diversity of entities at the quantum level is only quantitative and logically follows from the impossibility of one and the same entity entering into an irreflexive relation with itself. Dieks argued that weakening discernibility physically grounds the numerical diversity of physical objects only if the irreflexive relations at stake are physically relevant \cite [p. 8]{DieksLubberdink22}. From his point of view, it is still unclear how mathematical relations between formal objects defined in factorization spaces can be interpreted as representing relations between physical objects. The weakening of the notion of discernibility implies what Dieks calls \textit{factorism}, an account of the labeling of vectors as referring to numerically diverse particles.  
\begin{quotation}
\noindent {\small ``Without the presupposition of factorism we would just be studying mathematically defined irreflexive relations between component spaces of the total Hilbert space, and could only conclude that these factor spaces are different. That conclusion, however, would be trivial [...]  although it is true that spin measurements on the two wings of an EPR-Bohm experiment result in opposite outcomes, it is not clear that this is translatable into a statement about a preexisting relation between objects. Indeed, the usual interpretation of the correlations predicted by the singlet state is that these correlations testify to the holistic character of the spin system and should not be explained by an appeal to relations between spin properties that exist independent of measurement.'' \cite[p.9]{DieksLubberdink22}}     
\end{quotation}

Advancing a step further, Dieks, together with various collaborators, developed what has come to be known in the literature as the Alternative View (AV), according to which quantum entities can be defined in a contextual manner —restricted to certain specific experimental circumstances— as emergent classical particles \cite{DieksLubberdink11}. Contrary to the RV, Dieks suggests that, in general, what is usually identified as distinct identical particles in the same physical state should be understood as a single object and not as a system whose parts are considered as individuals. Arguing that the standard interpretation is problematic, Dieks proposes an alternative account of the standard formalism. As mentioned above, according to the Dutch physicist and philosopher, the problem with the usual interpretation of quantum theory involves the orthodox reading of the symmetrization postulate and what is critically known as factorism. According to quantum theory, states of identical particles must be either symmetric (in bossonic systems) or anti-symmetric (in fermionic systems). Thus, given a typical two-fermion state:

\[
|\Phi \rangle = \frac{1}{\sqrt{2}}  \{|\psi \rangle_1 |\phi \rangle_2 - | \phi \rangle_1 |\psi \rangle_2\}
\] \\
that is an element of the tensor product of a Hilbert space $\mathcal{H}_1 \otimes \mathcal{H}_2$ the labels \textit{1} and \textit{2}, that refer to the one-particle Hilbert spaces $\mathcal{H}_1$ and $\mathcal{H}_2$, distinguish two isomorphic copies of a Hilbert space $\mathcal{H}$. The usual orthodox interpretation is that states of this kind represents two individual particles, each associated with exactly one of the two labeled one-particle factor spaces. Dieks argues that this way of understanding the formalism is based on the idea that the labels added to the wave-functions individualize them as physical entities. However, the correlation  —which is, in classical physics, the basis for the use of particle labels— between one-particle state spaces and uniquely individuating particle states, i.e, that there is one identifying state for each particle, represented in its unique one-particle state space, is broken in QM \cite{Dieks23, DieksLubberdink22}. According to Dieks, the orthodox approach to quantum formalism implies that, in the quantum context, all particles of the same type have the same physical properties:
\begin{quotation}
\noindent {\small ``As the symmetrization postulates apply universally and globally to all particles of any given kind, for example all electrons in the universe, `factorists' must hold that each single electron is equally present at all positions in the universe at which there is `electron presence'. For example, according to factorism it does not make sense to speak about the specific electrons in my body, since all electrons in the universe are equally (partly) present there. This result is not restricted to localization but holds in the same way for whatever physical particle property one may think of. All electrons share all their physical properties and are therefore mutually indistinguishable in all respects. The very same holds for all protons, neutrons and other particles of the same kind in the universe.'' \cite[p. 6]{DieksLubberdink22}}     
\end{quotation}

Indeed, the orthodox view presupposes that quantum systems of the type described represent two individual particles associated with the formal labels  $\mathcal{H}_1$ and $\mathcal{H}_2$, even if, like in the RV, these are later characterized as non-individuals. Thus, Dieks asks in what sense these particles are significantly identified by formal labels if, in principle, they are permutable in the formalism. Furthermore, how can they be called particles if they are indistinguishable?
\begin{quotation}
\noindent {\small ``Particle indistinguishability is irreconcilable with the everyday meaning of “particle”, and also with how this term is used in the practice of physics. Moreover, it is a consequence of the standard view that identical quantum particles remain indistinguishable even in the classical limit, which makes a smooth transition to the classical particle concept impossible.'' \cite[p. 1]{DieksLubberdink22}}     
\end{quotation}

Contrary to the RV, Dieks argues that “quantum particles” are {\it emergent} and {\it contextual} entities in the sense that, at the fundamental level, the world is not made of particles, but once quantum particles “emerge” they are —in certain experimental contexts— distinguishable and possess identity, so that they can be dealt with using ordinary formal apparatuses \cite [p. 11] {Dieks23}. The idea, based on the theoretical framework of \cite{Ghirardi2002}, is to consider quantum particles, not as associated with factorization space labels, but with {\it one-particle states} that occur throughout the {\it N}-particle state, identifying —as occurs in practice— particles and physical states. Thus, in the AV, particles are identified with mutually orthogonal states of a single particle in a many-particle system, rather with factorial labels. The idea is that given a set of observables that are always symmetric in the labels of the factorial spaces, the properties of the candidate particles to be individualized must be represented by symmetric projection operators. Given operators of the type
\[
{\it P}_{1} \otimes {\it I}_{2} + {\it I}_{1} \otimes {\it P}_{2} - {\it P}_{1} \otimes {\it P}_{2}
\]
where P represents the projection operator to be used in the case of a one-particle system in which there is only one factorial space. The idea proposed by the AV is to associate the state of a particle onto which {\it P} is projected with a subsystem of a many-particle system consisting of multiple states. {\it N} pure states of a particle are associated with an antisymmetric state of {\it N} particles if the total state can be obtained from the symmetrization or anti-symmetrization of a product state of dimension {\it N}. In this reading of the quantum formalism, the subsystems are \textit{absolutely distinguishable} in the sense that they are differentiated on the basis of monadic “physical properties” (see \cite{DieksLubberdink22}). This characterization of quantum particles by virtue of arbitrary orthogonal states intends to provide a general picture of the individuality of particles, however, the connection between this way of understanding the formal apparatus and the field of phenomena of the theory must be explained. According to AV, this connection depends on the type of measurement: the particularized character of the system will manifest itself in some measurements and not in others \cite{Dieks2020}. In practice, the measurement and the interaction it entails will have a local character —relative to each experiment in question— so that the scenarios in which physical states are “spatially localized” are the states relevant to an interpretation of the notion of particle as advocated by the AV. In cases where a quantum system cannot be characterized according to the classic definition of individual, the authors suggest that a quantum system of, for example, bosons, can be better described as a quantum field. This approach to quantum formalism would be better suited, according to proponents of the AV, to the “actual practice of physics” because in it “particles” are not defined by reference to formal labels but by observed physical quantities.

In his reply, Krause {\cite{Krause25b} argues that the critics against factorism are unsuitable. According to the proponent of RV, formal labels are otiose in the sense that particles are permutable without changes in the state count, but this does not imply that reference is made to individual particles in the classical sense of individuality. Therefore, formal labels do not provide identity to any particle. Although quantum particles are indistinguishable in the formalism, the non-individual approach does not assume that there are individual particles in a quantum system in the sense that it does not attribute (classic) identity to them. Additionally, Krause asserts that the requirement that the properties or relations relevant to characterizing discernibility be \textit{physical}, that is, measurable and not logical, is problematic. According to him, no one in the discussion has managed to provide a rigorous definition of such a criterion. However, and despite the many differences, Krause argues that his approach and Dieks' are not conflicting interpretations but \textit{complementary} ones. While the non-individuals approach assumes that quantum entities can be understood as non-individuals that fail to obey classical identity and provides a formal apparatus for modeling non-individuality, Dieks' pragmatic approach is better suited to situations in which quantum particles can be located and have —momentarily— identity, so that it is possible to reason as if quantum particles had individuality. In this sense, Krause concedes the validity of the pragmatic approach for practical purposes (FAPP):
\begin{quotation}
\noindent {\small ``The AV is useful {\it for all physical purposes} (FAPP, as Bell used to say), and I think that it can be seen as complementary to the RV, which is more accurate from the logical point of view, once one accepts that quantum entities can be seen as non-individuals, something that looks perfectly possible.'' \cite[p.43]{Krause25b}}     
\end{quotation}

Let us now discuss, in the following section, a much deeper connection between the RV and the AV, which underlies the debate between these apparently rival approaches.

\section{The Anti-Realist Methodology Uncovered}

At first sight it might seem that Krause and Dieks, as main proponents of the RV and AV respectively, have radically opposing positions when discussing about quantum individuality (see \cite{Dieks25, Krause25b}). Indeed, to think that quantum particles are individuals or non-individuals might seem, in principle, antagonistic viewpoints. However, from a different perspective, this debate about quantum individuality can be also seen as being part of the same anti-realist methodology ---established in physics by Bohr and Dirac during the early 1930s--- which instead of critically addressing the reference of QM to particles has ---on the very contrary--- reinforced an unjustified self-contradictory atomist discourse that continues ---still today--- to be applied within contemporary physical and philosophical research. In this respect, it should be stressed that the reference to ``microscopic particles'' was built following an (anti-realist) empiricist methodology grounded on the distinction between what is observable and what is not. 

According to contemporary physics and philosophy, knowledge about the ``external world'' is {\it given} to us, humans, through the perception of individual observable ``macroscopic entities'' such as tables, chairs and dogs. These entities present themselves to us within experience, when we open our eyes, independently of any theoretical representation. As stressed by Tim Maudlin: 
\begin{quotation} 
\noindent {\small ``Any empirical  science has to start from what the philosopher Wilfred Sellars called `the manifest image of the world'; that is, the world as it presents to me when I open my eyes. And of course we know that some of those appearances can be deceptive or misleading ---you know, a straw in water looks bent but it really isn't and so on ...--- but you have nowhere else to begin but with the manifest image, and then you try and produce theories that would explain it or account for it.'' \cite{Maudlin19b}} 
\end{quotation}
Alisa Ney and David Albert explain this fundamental idea:
\begin{quotation}
\noindent {\small ``Any fundamental physical theory is supposed to account for the world around us (the manifest image), which appears to be constituted by three-dimensional macroscopic objects with definite properties. To accomplish that, the theory will be about a given primitive ontology: entities living in three-dimensional space or in space-time. They are the fundamental building blocks of everything else, and their histories through time provide a picture of the world according to the theory (the scientific image).'' \cite[p. 60]{NeyAlbert13}} 
\end{quotation}  

According to this way of thinking, theories and models are developed by scientists, firstly, in order to account for the individual entities we observe when we open our eyes. Accordingly, classical mechanics describes the motion of tables and chairs within space and time ---as part of our manifest image--- through an abstract mathematical {\it idealization} provided in terms of `rigid bodies'. This modelization implies an essential gap between what is {\it presented} (the target system) and its {\it re-presentation} (the rigid body). However, according to philosophers of science, the real problem arises when we consider `unobservable entities' and scientists are then forced to create less intuitive images that go beyond our ``common sense'', like quarks, black holes, dark matter, multiverses, etc. Thus, while classical mechanics can be justified as being part of the manifest image, the scientific image ---due to its departure from what is actually observable--- advances dangerously into ``ungrounded metaphysical illusions''.

In order to justify the continuity between classical and QM, both Bohr and Dirac would interpret these theories in terms of ``scales''. For example, as claimed in \cite[p. 1]{Dirac74}: ``it has been found possible to set up a new scheme, called quantum mechanics, which is more suitable for the description of phenomena on the atomic scale.'' Thus, regardless of the complete lack of any theoretical or experimental justification, it was simply assumed right from the start  that the phenomena described by the theory of quanta referred to ``small particles''. This idea, went completely against the historical development of the theory by Heisenberg who, in order to create the matrix formulation of QM, had abandoned Bohr's model of the atom and its need to represent the trajectories of unseen electrons.  Instead of a continuous development Heisenberg was proposing the need of radical conceptual jump: 
\begin{quotation}
\noindent {\small ``[...] the transition in science from previously investigated fields of experience to new ones will never consist simply of the application of already known laws to these new fields. On the contrary, a really new field of experience will always lead to the crystallization of a new system of scientific concepts and laws.''  \cite{Bokulich04}} 
\end{quotation}  
Of course, this understanding of physics confronted the Bohrian pragmatic approach ---followed also by Dirac\footnote{Heisenberg \cite[p. 101]{Heis71} describes Dirac as someone who ``felt that the development of our science was a more or less continuous process.  For once you use the pragmatic approach, you are bound to consider the progress of science as a continuous and never-ending process.'' According to Heisenberg \cite[p. 98]{Heis71}, ``[the] idea of continuous progress as we know it from engineering would weaken, or rather soften, physics to such an extent that we could hardly continue to call it an exact science.''}--- according to which QM had been derived as a ``continuous generalization of classical mechanics'' \cite{Bokulich04, Bokulich05}. But quite regardless of the fact the theory of quanta did not provide any meaningful representation of the notion of ``quantum particle'', the orthodoxy imposed by Bohr's doctrine of classical concepts ---justified through his principles of complementary and correspondence--- would soon become mainstream within the academic world. It is in this context that the contemporary debate about quantum individuality ---addressed mainly within philosophical journals--- has been set by presupposing right from the start that the target systems of the standard quantum model are ``microscopic particles''. The notion of ``quantum particle'' ---which implies within QM a self-contradictory narrative full of gaps and problems--- is not derived from the mathematical formalism or experimental findings but, instead, dogmatically imposed. The problem becomes then: how to account for these ---pre-theoretically presupposed--- microscopic entities? As stressed by Krause:
\begin{quotation}
\noindent {\small ``We have seen that contrary to what Ot\'avio Bueno, Francisco Berto and others say, something endorsed by Dieks et al., the notion of identity is not `essential' for the meaning of the concept of an entity. By an entity, we understand everything that can be referred to by a suitable language either by a proper name or by some description. An electron is an entity, and so are all quantum `particles'.''  \cite{Krause25b}} 
\end{quotation}

This debate implies an ontological quest whose goal is explained in the Oxford Dictionary in Philosophy \cite{OxfordDictionary}: ``Ontology is the study or concern about what kinds of things exist ---what entities or `things' there are in the universe.'' Thus, right from the start, the problem is to explain what is the real nature of quantum particles. The mainstream approach to solve this ontological problem implies the addition of an interpretational layer which is then understood as providing a {\it true} description of reality-itself. At this point, the realist is portrayed as a naive {\it believer} in his own made-up fictional illusions. Obviously, this position ---which supposedly captures realism--- can be easily criticized within the anti-realist scheme for there is no way ---for the scientific realist--- to justify why she would prefer one particular model and interpretation between the many available alternatives. Indeed, a particular chosen model plus interpretation will be just one possible combination between the huge cart of models and interpretations that can be consumed within philosophical journals \cite{Arroyo20}. This is the well known problem of {\it underdetermination} which has ---at least--- two layers. In the first layer, which is part of contemporary physics, it is commonly accepted that there will exist different frameworks, theories and models capable of providing different inconsistent or complementary accounts of the same `target system'. As explained by Ian Hacking: 
\begin{quotation}
\noindent {\small ``Various properties are confidently ascribed to electrons, but most of the confident properties are expressed in numerous different theories or models about which an experimenter can be rather agnostic. Even people in a team, who work on different parts of the same large experiment, may hold different mutually incompatible accounts of electrons. That is because different parts of the experiment will make different uses of electrons. Models good for calculations on one aspect of electrons will be poor for others. [...] There are lot of theories, models, approximations, pictures, formalisms, methods and so forth involving electrons, but there is no reason to suppose that the intersection of these is a theory at all.''  \cite[pp. 263-264]{Hacking83}} 
\end{quotation}

However, even if one would be able to choose one single model between the many, in the second interpretational layer, discussed in the philosophical arena, we are stuck with an analogous problem. As stressed by David Mermin \cite[p. 8]{Mermin12}: ``[Q]uantum theory is the most useful and powerful theory physicists have ever devised. Yet today, nearly 90 years after its formulation, disagreement about the meaning of the theory is stronger than ever. New interpretations appear every day. None ever disappear.'' This situation has lead the field into what \'Adan Cabello has termed an ``interpretational map of madness'' \cite{Cabello17}. Indeed, this permissive methodology that guides the construction of algorithmic models and fictional narratives has produced the complete fragmentation of knowledge and understanding. This might seem like a dead-end. But it is at this point that the constructive empiricist enters the scene in order to save the situation. By producing a dissection of {\it truth} in ontological and epistemological terms, she argues that even though a specific interpretation {\it could} be true and actually describe ``the stuff the world is made off'' (ontological truth); since we {\it cannot know} which {\it is} the true model and interpretation (i.e., epistemic truth) it might be preferable to remain simply ``agnostic''. This is, in fact, the stance made famous by ---the self-avowed anti-realist--- Bas van Fraassen: we should remain agnostic about the unobservable aspects of scientific theories, while accepting their empirical adequacy (i.e., their truth regarding observable phenomena). Even though interpretations might provide a representation of ``how the world {\it could be} the way that the theory says it is'' \cite[p. 337]{VF91}, they are incapable to say {\it which is} actually the case ---between the many possibilities. This means we can believe a theory is empirically adequate (i.e., its claims about observable things are true) without committing to belief in the existence or truth of its unobservable entities or mechanisms. This (anti-realist) position is completely widespread and shared by all researchers within this particular field of debate (e.g., Krause, Arenhart, Arroyo, Bueno, Dieks, Lubberdink, etc.). So, even though these researchers will fight for their chosen models and interpretations, none of them will actually claim that their model and interpretation is actually the true one.

What is thus essential to recognize is that, in all these cases, the {\it moments of unity} discussed by theories and models are presupposed even before any theoretical representation is being discussed. The fact there is no way to justify any true relation between the presupposed `target systems' (i.e, the ``quantum particles'') and the models plus interpretations ends up then reinforcing the acceptance of a multiplicity of a-systematic re-presentations justified ---in the end--- in purely pragmatic terms. This situation, which naturally leads to an extreme form of fragmentation (of models and interpretations), ends up justifying the anti-realist skepticism regarding theoretical representation itself. The underlying account is that: even though theories might provide a {\it true} (in a correspondent sense) access to ``the furniture of the world'', the problem is that {\it we cannot know}. True knowledge is simply unreachable and in the end, regardless of the many narratives we might posses, we should content ourselves with an instrumentalist account of phenomena. 

To sum up, we can list the set of presuppositions that have been implicitly assumed, not only by contemporary physicists but also by philosophers, as the main standpoints of empirical research: 
\begin{itemize}
\item  Observations, assumed as unproblematic {\it givens} of experience, are the basic standpoint of ``empirical science'' in general and of physical theories in particular. 
\item Physical theories are mathematical models with {\it correspondence rules} that allow us to make predictions about observational events. 
\item Conceptual narratives (i.e., interpretations) can be added to empirical models in order to {\it describe} how the world could possible be according to the theory (at least approximately).
\end{itemize}

Now, while in the macroscopic realm the correlation between classical mechanics and its target systems (e.g. tables, chairs and dogs) is justified by the match between what we observe and the representation provided by Newtonian mechanics in terms of tridimensional objects; in the microscopic realm things become much more complicated. Not only the target systems (e.g., electrons, protons, etc.) are un-observable but, even worse, the mathematical representation does not convey the expectations implied by any classical criteria. In this respect, there are many problems that arise when attempting to apply a classical account of individuality within QM:
\begin{enumerate}
\item[I.] The Kochen-Specker theorem demonstrates there is no consistent global binary valuation for the {\it projection operators} (representing properties) that pertain to the same quantum system \cite{deRondeMassri16, deRondeMassri21a, KS}. 
\item[II.] Quantum superpositions, at least before measurement collapses or projections, seem to describe systems in terms of contradictory properties \cite{daCostadeRonde13, Schr35}.
\item[III.] Even though it is commonly presupposed, the discreteness of the quantum postulate as well as the  multi-dimensionality of any vectorial representation preclude a spatial account of QM which is, of course, a pre-requisite to talk about ``microscopic scales'' \cite{Clauser03, Schr35}.\footnote{This essential point was stressed by Schr\"odinger himself and later on by David Bohm. As explained by John Clauser \cite[p. 29]{Clauser03}: ``Since quantum mechanics is formulated in a configuration space, it then provides no physical-space model to consult, whereupon it is then difficult to visualize what might actually be happening, especially in a two- (or more-) particle system.''} 
\item[IV.] The standard relativist definition of {\it quantum state} is an essentially inconsistent notion \cite{deRonde25b, deRondeMassri22a}. 
\end{enumerate} 

Now, taking for granted that QM talks about ``microscopic particles'', physicists and philosophers have learned to reinterpret these failures, not as clear expositions of the fact that the notion of particle is simply incompatible withe the mathematical formalism of the theory but, instead, as ways to characterize ``quantum particles'' either in purely pragmatic terms, as related to measurement outcomes, or in purely negative terms, namely, in terms of {\it what they are not}: in terms of their non-individuality, non-locality, non-identity, non-separability, etc. In the following section we propose to follow a different methodology for physics; one that was actually defended a century ago ---between many others--- by Einstein, Schr\"odinger, Heisenberg and Pauli.   


\section{Physical Individuals as Formal-Conceptual Machineries}

Contrary to the contemporary empiricist approach to scientific knowledge as grounded on the observation of individual entities {\it given} to us, humans, through our perception of the ``exterior world''; physics, as an attempt to account for {\it physis} (reality), was since its origin determined by the problem to account for objects of knowledge within change and becoming. The origin of physics goes back to the Ancient Greeks and the problem of movement, namely, given change and becoming, how to account for the multiplicity of experience in terms of {\it sameness}? Physics, as grounded on the notion of {\it  physis}, was never linked to the observation of macroscopic entities but to the systematic theoretical (formal and conceptual) construction of what could be considered to be {\it the same} within {\it change}. As explained by a young Wolfgang Pauli: 
\begin{quotation}
\noindent {\small  ``[...] knowledge cannot be gained by understanding an isolated phenomenon or a single group of phenomena, even if one discovers some order in them. It comes from the recognition that a wealth of experiential facts are interconnected and can therefore be reduced to a common principle. [...] `Understanding' probably means nothing more than having whatever ideas and concepts are needed to recognize that a great many different phenomena are part of coherent whole. Our mind becomes less puzzled once we have recognized that a special, apparently confused situation is merely a special case of something wider, that as a result it can be formulated much more simply. The reduction of a colorful variety of phenomena to a general and simple principle, or, as the Greeks would have put it, the reduction of the many to the one, is precisely what we mean by `understanding'. The ability to predict is often the consequence of understanding, of having the right concepts, but is not identical with `understanding'.'' \cite[p. 63]{Heis71}}
\end{quotation}

Contrary to antirealism which takes observation to be the standpoint of science,\footnote{This empiricist standpoint has been repeatedly confused with realism within the contemporary literature. Paul Dicken \cite[p. 66]{Dicken16}, for example, describes scientific realism as ``a kind of naive realism'' that ``might even seem like nothing more than just good common sense.'' Ronald Giere \cite[p. 4]{Giere06} writes: ``Everyone starts out a common-sense realist [...] For most people most of the time, common-sense realism works just fine. The realism of scientists may be thought of as a more sophisticated version of common-sense realism.''} realism has remained always suspicious of perception, centering its praxis in the production of formal conceptual machineries capable to provide a meaningful account of experience. This theoretical praxis is clear in Einstein's thought who argued explicitly: 
\begin{quotation}
\noindent {\small``From Hume Kant had learned that there are concepts (as, for example, that of causal connection), which play a dominating role in our thinking, and which, nevertheless, can not be deduced by means of a logical process from the empirically given (a fact which several empiricists recognize, it is true, but seem always again to forget). What justifies the use of such concepts? Suppose he had replied in this sense: Thinking is necessary in order to understand the empirically given, {\it and concepts and `categories' are necessary as indispensable elements of thinking.}'' \cite[p. 678]{Einstein65} (emphasis in the original)}
\end{quotation} 

It might be noticed that the conceptual development of a {\it moment of unity} should not be confused with the theory-ladenness of observation famously discussed ---between many others--- by Norwood Hanson, Thomas Kuhn and Paul Feyerabend. While in the latter case, the idea is that the target object even though \emph{given} within experience can be interpreted and shaped by different theories and conceptual frameworks; in the case we are discussing here we argue it is the very notion of object that must be explicitly defined in categorical terms. This means that in physics, conceptual systems do not provide interpretations of an observed object but ---as Einstein frequently emphasized--- impose the very conditions that determine what can be considered to be {\it the same} within observation. 
\begin{itemize}
\item  The understanding of what is observed presupposes adequate concepts that are able to constrain experience in terms of specific {\it moments of unity}. 
\item Theories are systems which, through the consistent and coherent interrelation between mathematical formalisms and conceptual schemes, are able to account in operational terms for specific fields of phenomena. 
\item A conceptual scheme is not just a narrative added to a mathematical model but ---instead--- a categorical system which allows to determine what is {\it the same} within change regardless of the empirical perspective.
\end{itemize}

Accordingly, scientific observation cannot be regarded in naive (pre-theoretical or pre-conceptual) terms as referring to {\it given} objects of experience. Instead, observation must be understood as a necessary consequence of the theory itself, in which case, it becomes essential to recognize the fundamental role of conceptual systems within theories. This is the reason why Einstein would remark to a young Heisenberg \cite{Heis71} that: ``It is only the theory which tells you what can be observed.'' Or, as expressed by Heisenberg himself: 
\begin{quote}
\noindent {\small ``The history of physics is not only a sequence of experimental discoveries and observations, followed by their mathematical description; it is also a history of concepts. For an understanding of the phenomena the first condition is the introduction of adequate concepts. Only with the help of correct concepts can we really know what has been observed.''   \cite[p. 264]{Heis73}}
\end{quote}
 

While in the case of anti-realism individual entities are conceived as {\it given} within experience ---and then interpreted in different inconsistent or complementary fashions---, in the realist case ``it only the theory which tells you what can be observed.'' Every conceptual system provides its own conditions for determining what is {\it the same}, and therefore an individual. While in classical mechanics it is the notion of `particle' which provides unity, in the case of electromagnetism it is the notion of `electromagnetic wave'. Physical individuality, in this latter case, must be regarded as a formal-conceptual machine created in order to provide unity to the multiplicity of experience. In this respect, the general principles that guide and constrain physical representation are {\it operational-invariance} and {\it conceptual-objectivity}. Thus, while anti-realism is grounded in a pre-theoretical experience ---presupposing a direct access to an ``exterior world'' composed of individual entities--- in the realist case physics has always been guided by the attempt to solve the problem of movement in theoretical terms, namely, providing a specific set of metaphysical principles capable to forge identity within difference. In this respect, we might recall that one of the first systematic solutions to the problem of movement would be developed by Aristotle himself who, in  {\it Metaphysics}, would craft an explicit definition of {\it actual entity} in terms of the principles of existence, non-contradiction and identity (see for a detailed discussion \cite{dDFInternet}). It is the establishment of the primacy of theoretical conditions over experience, and on the contrary, of observation over theoretical production, what distinguishes the realist theoretical path from the anti-realist empiricist one. It is by following the Greek realist methodology, seeking for systematic theoretical unity, that Galileo would advance the scientific program one step further by bringing into unity the different frame dependent representations ---through an {\it invariant transformation}. Classical mechanics would then unite not only the terrestrial and celestial realms, but ---even more importantly--- the seemingly different phenomena observed from distinct viewpoints or reference frames. And it is exactly this same praxis that would allow Maxwell to unify magnetic and electric phenomena in terms of the same entity. In turn, also Einstein would stay close to the {\it relativity principle} in his special theory of relativity in order produce a consistent invariant account of phenomena abandoning the spatiotemporal substantialist representation that had been inherited from classical mechanics. Thus, the question raises: is there an equivalent path for QM? Can we produce an invariant-objective concept of individual for the theory of quanta?

\section{A Relational Account of Quantum Relative Individuals}

In a new line of research called the Logos Categorical Approach to QM it has been argued that when following the guide of operational-invariance it is possible to escape the contemporary maze created by the inconsistent axiomatic formulation of the theory interpreted also in an inconsistent fashion in terms of quantum particles, measurements, collapses, preferred bases, etc.. This is indeed what is known today by physicists as the ``Standard'' formulation of QM which ---as famously stated by Richard Feynman---  no one really understands. However, by staying close to the original methodology of physics, it is easy to see that Heisenberg's matrix mechanics (with its original reference to intensive values) is an already consistent operational-invariant mathematical formalism that can easily escape not only the relativist contextual definition of quantum state, but also the need of irrepresentable ``quantum jumps'' and ``collapses'' \cite{deRonde25b, deRondeMassri20, deRondeMassri22a}. As it has been demonstrated \cite{deRondeMassri16}, referring to intensities instead of binary outcomes, it is possible to bypass Kochen-Specker contextuality restoring an invariant account of {\it projection operators} and thus a consistent global (intensive) valuation. Furthermore, following this line of research, it is possible to produce a definition of a quantum individual which, even though becomes relative to reference frames, is nonetheless both invariant and objective \cite{deRondeFMMassri25, deRondeMassri23}. In order to advance let us recall a series of definitions given in \cite{deRondeFMMassri24}.  

We begin with the definition of a (simple) \emph{graph} as a pair $G = (V, E)$, where $V$ is a set whose elements are called vertices (or nodes), and $E$ is a set of unordered pairs $\{v,w\}$ of vertices, whose elements are called edges. While each {\it vertex} is related to the mathematical notion of {\it projector operator} and to the physical concept of  {\it power of action}, each {\it edge} is linked to the mathematical concept of {\it commutation} and the physical compatibility of powers within an experimental arrangement. 

\begin{definition} {\bf Graph of powers:} Given a Hilbert space $H$, the graph of powers $G(H)$ is defined such that the vertices are the projectors on $H$ (called \emph{powers}), and an edge exists between projectors $P_1$ and $P_2$ if they commute.
\end{definition} 

\noindent It is these powers, in their multiplicity and their relationships, that allow us to define an \emph{Intensive State of Affairs} (ISA) ---in contrast to an {\it Actual (or Binary) State of Affairs} (ASA). But first, we need to formalize the notion of intensity (or potentia). The assignment of intensities is called {\it Global Intensive Valuation} (GIV).

\begin{definition} {\bf Global Intensive Valuation:}  A Global Intensive Valuation is a map from $G(H)$ to the interval $[0,1]$.
\end{definition}

\noindent Clearly, not all GIV's are compatible or consistent with the relations between powers. We will focus on those that define an ISA as follows:

\begin{definition} {\bf Intensive State of Affairs:} Let $H$ be a Hilbert space of infinite dimension. An \emph{Intensive State of Affairs} is a GIV $\Psi: G(H)\to[0,1]$ from the graph of powers $G(H)$
such that $\Psi(I)=1$ and 
\[
\Psi(\sum_{i=1}^{\infty} P_i)=\sum_{i=1}^\infty \Psi(P_i)
\]
for any piecewise orthogonal operator $\{P_i\}_{i=1}^{\infty}$. The numbers $\Psi(P) \in [0,1]$ are called {\it intensities} or {\it potentia} and the vertices $P$ are called \emph{powers of action}. Taking into consideration the ISAs, it is then possible to advance towards a consistent GIV which can bypass the contextuality expressed by the Kochen-Specker Theorem \cite{deRondeMassri21a, KS}. 
\end{definition} 
At this point, it is necessary to say that a fundamental characteristic of any conceptual representation of a physical theory is its operationality. That is, the ability to relate physical concepts to what is actually observed in experiments. As stressed by Einstein \cite[p. 26]{Einstein20}, a physical concept lacks value if we are unable to connect it with an experimental corroboration. 
\begin{definition} {\bf Quantum Laboratory:} We use the term \emph{quantum laboratory} (or quantum lab or Q-Lab) as the operational concept of an ISA. 
\end{definition} 
\begin{definition}{\bf Screen and Detector:} A \emph{screen} with $n$ places for $n$ detectors corresponds to the vector space $\mathbb{C}^n$. Choosing a basis, say $\{|1\rangle,\dots,|n\rangle\}$, is the same as choosing a specific set of $n$ {\it detectors}. A \emph{factorization} $\mathbb{C}^{i_1}\otimes\dots \otimes\mathbb{C}^{i_n}$ is the specific number $n$ of screens, where the screen number $k$ has $i_k$ places for detectors, $k=1,\dots,n$. Choosing a \emph{basis} in each factor corresponds to choosing the specific detectors; for instance $|\uparrow\rangle, |\downarrow\rangle$. After choosing  a basis in each factor, we get a basis of the factorization $\mathbb{C}^{i_1}\otimes\dots \otimes\mathbb{C}^{i_n}$
that we denote as
\[
\{ |k_1\dots k_n\rangle \}_{1\le k_j\le i_j}.
\]
\end{definition} 

\begin{definition}{\bf Power of action:} The \emph{ basis element} $|k_1\dots k_n\rangle$ determines the \emph{ projector}  $|k_1\dots k_n\rangle \langle k_1\dots k_n|$ which is the formal-invariant counterpart of the objective physical concept called \emph{power of action} (or simply \emph{power}) that produces a global effect in the $k_1$ detector of the screen $1$,  in the $k_2$ detector of the screen $2$ and so on until the $k_n$ detector of the screen $n$. Let us stress the fact that this effectuation does not allow an explanation in terms of particles within classical space and time. Instead, this is explained as a characteristic feature of powers. In general, any given power will produce an intensive multi-screen non-local effect. 
\end{definition} 

\begin{definition} {\bf Experimental Arrangement:} Given an ISA, $\Psi$, a factorization $\mathbb{C}^{i_1}\otimes\dots \otimes\mathbb{C}^{i_n}$ and a basis $B=\{|k_1\dots k_n\rangle\}$ of cardinality $N=i_1\dots i_n$, we define an \emph{ experimental arrangement} denoted $EA_{\Psi,B}^{N, i_1\dots i_n}$, as a specific choice of screens with detectors together with the potentia of each power, that is,
\[
EA_{\Psi,B}^{N,i_1\dots i_n}= \sum_{k_1,k_1'=1}^{i_1}\dots \sum_{k_n, k_n'=1}^{i_n} 
\alpha_{k_1,\dots,k_n}^{k_1',\dots,k_n'} | k_1 \dots k_n \rangle \langle k_1' \dots k_n'|.
\]
Where the number $N$ is the cardinal of $B$ and is called the \emph{degree of complexity} (or simply \emph{degree}) of the experimental arrangement. 
\end{definition} 
\begin{definition}{\bf Potentia:} The number that accompanies the power $|k_1\dots k_n\rangle \langle k_1\dots k_n|$ is its \emph{potentia} (or intensity) and the basis $B$ determines the powers defined by the specific choice of screens and detectors. 
\end{definition} 

Assume that, in a Q-Lab,  we want to change or modify an experimental setup by changing the number of screens and detectors. In this case, there are two important theorems (derived in \cite{deRondeMassri23}) that allow us to consider the relations between powers, intensities, experimental arrangements and quantum labs. If the number of powers (i.e., the degree of complexity) remains the same after the rearrangement, then the {\it Basis Invariance Theorem} tell us that the new experimental arrangement is equivalent to the previous one, but if the complexity of the new experimental arrangement drops, then the {\it Factorization Invariance Theorem} tell us that all the knowledge in the new experimental arrangement was already contained in the previous one (see for a detailed analysis \cite{deRondeFMMassri24b, deRondeFMMassri25}). 

\begin{theo}{\sc (Basis Invariance Theorem)}
Given a specific Q-Lab $\Psi$, all experimental arrangements of the same complexity, $EA_{\Psi}^N$, are equivalent independently of the basis. 
\end{theo}

\begin{theo} {\sc (Factorization Invariance Theorem)}
The experiments performed within a $EA_{\Psi}^N$ can also be performed with an experimental arrangement of higher complexity N+M, $EA_{\Psi}^{N+M}$, that can be produced within the same Q-Lab $\Psi$.  
\end{theo}
\noindent Following the {\it Factorization Invariance Theorem}, we can thus say that a more complex experimental arrangement (one which considers more powers) includes less complex experimental arrangements. Meaning that all EAs can be completely deduced from any EA that is more complex. This also implies that the more intensive powers we consider ---that is, the more complex the individual---, the more knowledge we will gain of the state of affairs. This is completely different from the orthodox case in which a pure state is supposed to provide maximal knowledge. In our case, the knowledge is directly related to the level of complexity considered, not to purity. In fact, a pure state of degree 1 represented by a one term {\it ket}, $| \psi \rangle$, will provide the minimal possible information about that particular state of affairs.

Given these theorems it is possible to present the following definition of a {\it quantum individual}: the specific multiplicity of powers that allows, in each case, to completely determine not only a physical situation but also all its possible transformations. Or, to put it more precisely: the minimum set of powers of action within a specific degree of complexity capable of deriving the totality of powers and potentia in that same degree (or less). 
\begin{definition}{\bf  Quantum Individual:}  A quantum individual is a set of powers of complexity $N$ capable to derive the totality of powers and their respective potentia in that specific degree (or less).
\end{definition}
Something characteristic of this relative individual is that, having an EA of a certain complexity, that is, a specific set of powers, we can deduce any other EA of that same degree of complexity (or less), determined by another set of powers. Thus, having a consistent set of powers with their potentia, we can deduce all the other powers (and their respective potentia) in that same degree of complexity. The fact that given a specific multiplicity of powers (a relative individual) we can deduce a different set of powers from the same state of affairs expresses the inherent connectedness we previously described, the intrinsic systematic and relational nature of this representation. The contrast with the atomist picture in this case is instructive, since, given a multiplicity of particles, if we know the value of the properties of some particle we cannot deduce the values of the properties of other particles.

\section{A Methodological Dispute: Bohr vs Einstein (and Schr\"odinger)}

In a nutshell, what we have discussed in this work encircles the methodology required by physics in order to make sense of a specific field of phenomena in terms of {\it sameness}. While the ``Received'' and ``Alternative'' views, following the standard approach to QM established by Bohr and Dirac, have taken as a natural standpoint the notion of `particle' in order to discuss the different models of the theory, the Logos Categorical Approach, following Einstein's lessons regarding the kernel role of operational-invariance and conceptual-objectivity within physical representation, proposes instead to replace this dogmatic standpoint by the operationally-invariant formulation of QM that was originally created by Heisenberg in order to consistently address intensive phenomena (not binary `clicks' in detectors) and from there on generate what Bohr had explicitly prohibited, namely, a new system of interrelated physical concepts capable to provide an objective, consistent and coherent, unified representation of quantum phenomena. We might thus argue that while the ``Received'' and ``Alternative'' views can be regarded as neo-Bohrian approaches to quantum individuality, the Logos Categorical Approach stands in line with the realist methodology of physics that was defended by Einstein and Schr\"odinger. As Einstein would write in a letter dated 22 December 1950 to his fellow companion: 
\begin{quotation}
\noindent {\small ``You are the only contemporary physicist, besides Laue, who sees that one cannot get around the assumption of reality ---if only one is honest. Most [physicists] simply do not see what sort of risky game they are playing with reality ---reality is something independent or what is experimentally established.''  \cite[p. 39]{ESPL}} 
\end{quotation}

Having drawn this line, it becomes important to address what we consider an inaccurate interpretation of Schr\"odinger's writings as provided by the self-proclaimed ``Received View''. According to our account of his work, Schr\"odinger did not argue, as implied by da Costa and Krause, that ``QM talks about quantum particles which must be non-individuals''; what he argued repeatedly was that the very notion of `particle' was simply incapable to address the mathematical formalism of the theory ---an idea shared also by Einstein. Against the standard paradoxical reference to microscopic entities established by Borh and Dirac, Schr\"odinger would stress the fact that the term ``atom'' had become ``a misnomer'' \cite[p. 183]{Schr50}. More explicitly, he argued that: ``modern atomic theory has been plunged into a crisis. There is no doubt that the simple particle theory is too na\"ive.'' \cite[p. 87]{Schrodinger14}. According to the Austrian physicist: ``We have taken over from previous theory the idea of a particle and all the technical language concerning it. This idea is inadequate. It constantly drives our mind to ask information which has obviously no significance'' \cite[p. 188]{Schr50}. As already noticed, Schr\"odinger \cite[p. 183]{Schr50} would then conclude that when considering the notion of particle it becomes impossible to sustain its identity: an ``elementary particle'', he says, ``is not an individual; it cannot be identified, it lacks `sameness'.'' This should not be read as implying the need to detach particles from their individuality or identity,  but ---instead--- as exposing the limits of the notion of particle itself in the context of QM. {\it If} (and only {\it if}) we accept the notion of `particle' as a necessary presuposition of QM ---as Bohr and Dirac did within their ``standard'' formulation--- then we reach an absurdity, an {\it oximoron}, namely, that particles have no individuality. To support our reading we might mention that Schr\"odinger had already applied this  {\it ad absurdum} strategy within his famous ``cat paradox'' where the notion of `cat' was used, not to argue in favor of the existence of half dead and half alive ``zombie cat'', but ---on the very  contrary--- to expose a radical limitation of the classical notion of object when related to the mathematical formalism of QM. As he would clearly explain: 
\begin{quotation}
\noindent {\small ``[...] if I wish to ascribe to the model at each moment a definite (merely not exactly known to me) state, or (which is the same) to {\it all} determining parts definite (merely not exactly known to me) numerical values, then there is no supposition as to these numerical values {\it to be imagined} that would not conflict with some portion of quantum theoretical assertions.'' \cite[p. 156]{Schr35}}
\end{quotation} 

To sum up, Schr\"odinger never argued in favor of the notion of particle. Instead, his writings clearly point in the direction of abandoning this classical notion in order to address QM consistently. And this is exactly the path taken by the Logos Categorical Approach. The problem is not how to change the notion of `particle' in order to make sense of the theory of quanta but, instead, to accept the limitations of the classical representation and move forward, producing a new objective-invariant definition of individual capable to account for quantum phenomena. This essential point was also stressed by Einstein:
\begin{quotation}
\noindent {\small ``Concepts that have proven useful in ordering things easily achieve such an authority over us that we forget their earthly origins and accept them as unalterable givens. Thus they come to be stamped as ‘necessities of thought,’ ‘a priori givens,’ etc. The path of scientific advance is often made impossible for a long time through such errors. For that reason, it is by no means an idle game if we become practiced in analyzing the long commonplace concepts and exhibiting those circumstances upon which their justification and usefulness depend, how they have grown up, individually, out of the givens of experience. By this means, their all-too-great authority will be broken. They will be removed if they cannot be properly legitimated, corrected if their correlation with given things be far too superfluous, replaced by others if a new system can be established that we prefer for whatever reason.'' \cite{Howard10}}
\end{quotation} 

Finally, let us remark that the RV's notion of quantum non-individual does not possess any explicit categorical definition. Instead, it is constructed through the extraction of one of the main constituent elements that characterizes the notion of `particle' itself. Thus, what we obtain is a negative definition which tells us what quantum non-individuals are not; evading in this way the answer to the question: what are quantum non-individuals? 

As we have seen above, the AV argues in a much more direct manner for the applicability of the classical notion of `particle' within QM. One of the main arguments presented by Dieks \cite[p. 278]{Dieks25} is that: ``There are countless [...] examples in the physics literature in which identical quantum particles are treated as individual entities.'' Given that physicists are the experts in their field and that they do talk explicitly  about ``particles'', philosophers should uncritically accept this mainstream discourse as a necessary standpoint. This argument from authority ({\it ad verecundiam}), apart from being invalid, seems to forget about the historical development of physics in the 20th century. It is well known that after the IIWW physics would become considered in a purely instrumentalist fashion. As described by Olival Freire Jr.: 
\begin{quotation}
\noindent {\small ``In the US, which after the Second World War became the central stage of research in physics in the West, the discussions about the interpretation of quantum mechanics had never been very popular. A common academic policy was to gather theoreticians and experimentalists in order to favour experiments and concrete applications, rather than abstract speculations (Schweber 1986). This practical attitude was further increased by the impressive development of physics between the 1930s and the 1950s, driven on the one hand by the need to apply the new quantum theory to a wide range of atomic and subatomic phenomena, and on the other hand by the pursuit of military goals. As pointed out by Kaiser (2002, pp. 154-156), `the pedagogical requirements entailed by the sudden exponential growth in graduate student numbers during the cold war reinforced a particular instrumentalist approach to physics'.'' \cite[pp. 77-78]{Freire15}} 
\end{quotation}
Due to this instrumentalist turn, QM is presently taught to students in physics as an algorithmic model capable to account for measurement outcomes. As explained by Tim Maudlin: 
\begin{quotation}
\noindent {\small ``What is presented in the average physics textbook, what students learn and researchers use, turns out not to be a precise physical theory at all. It is rather a very effective and accurate recipe for making certain sorts of predictions. What physics students learn is how to use the recipe. For all practical purposes, when designing microchips and predicting the outcomes of experiments, this ability suffices. But if a physics student happens to be unsatisfied with just learning these mathematical techniques for making predictions and asks instead what the theory claims about the physical world, she or he is likely to be met with a canonical response: Shut up and calculate!'' \cite[pp. 2-3]{Maudlin19}} 
\end{quotation} 

Today, the question about the reference of the theory of quanta with respect to reality is simply not considered as part of physics. As recently described by Sean Carroll \cite{Carroll20}: ``Many people are bothered when they are students and they first hear [about SQM]. And when they ask questions they are told to shut up. And if they keep asking they are asked to leave the field of physics...'' That is in fact the main reason why ``philosophy of QM'' was established as a new discipline during the 1980s. It is true ---as mentioned by Dieks--- that contemporary physicists do supplement their computations with the well known microscopic narrative that was established by Bohr and Dirac almost a century ago. Indeed, it is very easy to find references to ``quantum particles'' within today's mainstream physical literature. But this does not mean that physicists use these notions consistently. On the contrary, it is full of examples where exactly the opposite takes place.\footnote{A good example is the field of quantum entanglement where the reference to particles, even though explicit in most papers, becomes immediately problematic when considered in relation to the standard mathematical formalism. See: \cite{deRondeFMMassri24b, deRondeMassri20}.} The truth is that, when pushed hard enough, physicists are ready to recognize that their discourse about quantum particles is ---at most--- ``just a way of talking''. As explained by Xiao-Gang Wen \cite{Wolchover20}, a theoretical physicist at the Massachusetts Institute of Technology: ``We say they are `fundamental'. But that’s just a [way to say] to students, `Don't ask! I don't know the answer. It's fundamental; don't ask anymore'.'' Given this situation, it seems to us that philosophers of QM should be much more critical regarding a discipline who trains their students even today to: ``Shut up and calculate!'' To repeat what physicists say in an un-critical fashion is not the way to address the problem but ---on the very contrary--- to reinforce it. 


The methodology of the AV is copied from Bohr: taking for granted his {\it doctrine of classical concepts} ---instead of creating the necessary concepts that could account in a consistent and coherent manner for quantum phenomena--- the AV proposes to restrict the context of inquiry, the experimental conditions, in order to apply a classical notion which is imposed as a necessary requirement to make sense of experience. An analogous exercise could be done with electromagnetism. It is true that in certain circumstances, under certain regimes, it could be possible to describe certain electromagnetic phenomena applying the notion of `particle' without ever mentioning `electromagnetic waves'. In very precise circumstances that could actually work out. But of course, when changing just a little bit the experimental context the description would immediately collapse. The reason is simple: particles cannot be used in general in order to explain electromagnetic phenomena. So what do we gain with this exercise of explaining phenomena with inadequate concepts? It seems that, apart from reassuring our prejudices regarding our understanding of reality, not very much. Something similar takes place when Dieks \cite{Dieks25} attempts to justify his approach in terms of decoherence: ``it is not assumed that particles are fundamental building blocks of the quantum world. Rather, particles are emergent entities [...]. Only if certain conditions are fulfilled (e.g., relating to decoherence or the classical limit) does a description in terms of particles become appropriate.'' Just like Bohr made use of his correspondence principle in order to justify his atomic model \cite{BokulichBokulich20}, Dieks mentions decoherence and the classical limit in order to justify his application of the notion of particle within QM. The problem ---once again--- is that it is well known in the philosophical literature that decoherence fails to solve the quantum to classical limit (or the measurement problem). As stated by Guido Bacciagaluppi \cite{Bacciagaluppi25}: ``Does modelling measurements including the decoherence interactions with the environment allow one to derive that measurements always have results? This is somewhat part of the ‘folklore’ of decoherence, but as pointed out by many physicists and philosophers alike [...], it is not the case.'' During the 1990s, mainly due to the work of philosophers, it was explicitly shown that decoherence was a precarious model full of technical and conceptual problems which was simply incapable to produce a consistent and coherent explanation of how quantum superpositions ended up turning into tables and chairs. As a strange way of recognizing this criticism, physicists even though finally accepted nothing had been solved, created a new type of ``solution'' called ``For All Practical Purposes'' ---in short, FAPP--- which, again, can be resumed in the following terms: ``It works, so shut up and calculate!''

\smallskip

Taking distance from these essentially neo-Bohrian approaches ---which are even regarded by Krasue as being complementary to each other (section 1)---, we propose to return to the original methodology of modern physics where the formal-conceptual construction of individuals was always considered as kernel to physical representation itself. And just like classical mechanics was not created by observing particles, the theory of electromagnetism was not created by observing electromagnetic waves. On the contrary, these concepts were built in terms of general metaphysical principles in order to account and unify a multiplicity of different phenomena. This is in fact, the same methodology which guided not only Einstein in the development of his special theory of relativity but also Heisenberg in his search for an operationally-invariant formalism that would capture the quantum intensities that were measured in the lab. Instead of staying close to the ``commonsensical'' spatiotemporal representation provided by Newton, Einstein would advance ---taking as a standpoint the {\it special principle of relativity} as well as {\it Lorentz' transformation}--- into a new conceptual system capable to secure the experimental finding of the invariance of the speed of light. The same applies to Heisenberg who, instead of discussing the problem of the trajectory of unseen electrons would advance beyond the Bohrian model in order to deliver a consistent operational-invariant account of quantum phenomena. This is exactly the methodology which the Logos Categorical Approach attempts to restore: stay close to the operational-invariance of intensive phenomena already encapsulated in Heisenberg's formalism, and advance towards new physical notions capable to explain in qualitative terms what is actually being observed according to the theory. The price to pay is to abandon not only the classical representation that Bohr wanted to retain at all costs but also the ontological account of physical individuals that is entrenched with it. Indeed, our analysis leads to the (objective) relativization of quantum individuality itself. This involves an interesting analogy: just like Einstein's theory involves a relativization of space and time, the theory of quanta implies the need to relativize the notion of individuality itself. 


To conclude, the problem of individuality in QM highlights the need to reconsider the methodology that was imposed to physics during the 1930s and continues, even today, to play an essential role within contemporary research. What is at stake here is the meaning of physics itself understood, either as a family of inconsistent algorithmic models supplemented by fictional narratives or as linked to the production of closed systematic theories capable to account in formal-conceptual terms, invariantly and objectively, consistently and coherently, for a specified field of phenomena. In particular, the theory of quanta points to the need of considering an essentially new experience, one that ---contrary to Bohr and neo-Bohrian approaches--- cannot be described with classical concepts.


\end{document}